\documentclass[aps,prb,twocolumn,showpacs,10pt,floatfix,longbibliography]{revtex4-2}

\usepackage{graphicx}
\usepackage{subfigure}
\usepackage{amsmath}
\usepackage{amsfonts}

\usepackage{mathrsfs}
\usepackage{bbm}
\usepackage{subfigure}
\usepackage{xcolor}

\usepackage[colorlinks=true, urlcolor=blue, linkcolor=blue, citecolor=blue]{hyperref}

\usepackage[T1]{fontenc}

\usepackage{siunitx}
\usepackage{bm}
\usepackage[version=4]{mhchem} 

\usepackage{stackengine}[2013-09-11]

\usepackage{dsfont}

\begin{document}

\title{Tunneling spin Hall effect induced by unconventional $p$-wave magnetism}

\author{W.~Zeng}
\email{zeng@ujs.edu.cn}
\affiliation{Department of Physics, Jiangsu University, Zhenjiang 212013, China}

\begin{abstract}
We propose a tunneling spin Hall effect in a normal metal/$p$-wave magnet/superconductor junction. It is found that the Andreev reflection in the normal lead is spin-dependent and exhibits strong asymmetry with respect to the transverse momentum, giving rise to a pure transverse spin Hall current with zero net charge. The transverse spin conductance is analytically derived using the nonequilibrium Green's function approach, revealing that the predicted spin Hall effect is governed by the direction of the Fermi surface splitting in the $p$-wave magnet. A finite transverse spin current with a large spin Hall angle arises when the line connecting the centers of the spin-split Fermi surfaces is perpendicular to the normal direction of the junction, which indicates a highly efficient charge-to-spin conversion, suggesting potential applications in spintronic devices.

\end{abstract}
\maketitle

\section{Introduction} 
\label{sec:int}

Unconventional magnets have recently emerged as a hot topic in condensed matter physics \cite{PhysRevX.14.011019,PhysRevLett.134.086701,PhysRevLett.133.236703,PhysRevB.111.144508}. Those exhibiting even-parity magnetic order (\textit{e.g.}, $d$-wave symmetry) typically preserve the inversion symmetry ($\mathcal{P}$) while breaking the time-reversal symmetry ($\mathcal{T}$), and are referred to as altermagnets. Several materials have been proposed as candidates for altermagnets, including $\ce{RuO2}$ \cite{PhysRevX.12.031042,PhysRevB.99.184432}, $\ce{Mn5Si3}$ \cite{PhysRevX.12.011028,reichlova2024observation}, and $\ce{MnTe}$ \cite{PhysRevB.107.L100418,PhysRevLett.132.036702}. On the other hand, the emergence of the odd-parity magnetic order with $p$-wave symmetry gives rise to the unconventional $p$-wave magnets \cite{PhysRevB.111.L220403,10.21468/SciPostPhys.18.6.178,PhysRevB.111.054501}, which are characterized by preserved $\mathcal{T}$ and broken $\mathcal{P}$. Such unconventional $p$-wave magnets have been proposed in materials like $\ce{Mn3GaN}$ and $\ce{CeNiAsO}$ \cite{PhysRevX.12.011028}. Both the $d$-wave altermagnets and the unconventional $p$-wave magnets feature the momentum-dependent nonrelativistic spin splitting, which can be significantly larger than that induced by spin-orbit interactions, thereby attracting considerable attention in the field of charge and spin transport \cite{PhysRevB.110.134437,nrk5-5zrj,PhysRevB.111.125119,PhysRevB.111.035132}. The tunneling magnetoresistance and spin transport in the normal metal/$p$-wave magnet junctions have recently been studied \cite{PhysRevLett.133.236703,PhysRevB.111.035404}. In addition, the interplay between unconventional magnets and superconductors in heterostructures is of particular interest, with several novel transport phenomena that have been reported, such as the orientation-dependent Andreev reflections in altermagnets \cite{PhysRevB.108.054511,PhysRevB.108.L060508}, enhanced crossed Andreev reflections in the junctions composed of $p$-wave magnets and spin-triplet superconductors \cite{PhysRevB.111.165413}, $\varphi$ Josephson junctions \cite{PhysRevLett.133.226002} and spin-dependent Andreev levels \cite{PhysRevB.111.064502,PhysRevB.108.075425}.

The spin Hall effect provides a fundamental route to pure spin current control and is central to the advancement of spintronics \cite{RevModPhys.87.1213,PhysRevLett.83.1834,PhysRevLett.96.106802,10.21468/SciPostPhys.18.5.146}. The conventional spin Hall effect can be generated via the intrinsic band structure effects or extrinsic impurity scattering. However, in ballistic heterojunctions, the spin-dependent skew tunneling serves as an alternative mechanism for generating a finite transverse spin current, which is termed the \textit{tunneling} spin Hall effect \cite{PhysRevLett.115.056602,10.1063/1.5051629,PhysRevLett.117.166806,PhysRevLett.110.247204}. Previous studies reported that the tunneling spin Hall effect occurs due to the symmetry mismatch in ferromagnetic tunnel junctions \cite{PhysRevLett.115.056602,10.1063/1.5051629}. Moreover, the Rashba and Dresselhaus spin-orbit coupling in the barrier of ferromagnet/superconductor junctions may also lead to a finite transverse spin current in superconducting tunnel junctions \cite{PhysRevB.100.060507,PhysRevB.111.054512}. Typically, the tunneling spin Hall effect is independent of the Berry curvature; however, it can still be realized through the Berry-curvature-related mechanism, such as the geometric phase-coherent mechanism \cite{RevModPhys.64.51,PhysRevB.111.075418,PhysRevB.87.165420}.

In this paper, we demonstrate that $p$-wave magnetism can also give rise to the tunneling spin Hall effect in superconducting tunnel junctions. We theoretically investigate the transverse charge and spin transport in the normal metal/$p$-wave magnet/superconductor junction. It is found that the $p$-wave magnetism may lead to the spin-dependent Andreev reflection in the normal lead. Notably, this spin-resolved Andreev reflection is strongly asymmetric with respect to the transverse momentum, resulting in spin-selective transverse transport, where the carriers with opposite spins propagate along opposite transverse directions of the tunnel junction, manifesting as the tunneling spin Hall effect. The transverse spin conductance is obtained using the nonequilibrium Green's function (NEGF) method and exhibits an orientation dependence on the direction of the Fermi surface splitting. A finite transverse spin current with a large spin Hall angle can be generated when the line connecting the centers of the spin-split Fermi surfaces is perpendicular to the normal direction of the junction, indicating a highly efficient charge-to-spin conversion. 

The remainder of this paper is organized as follows. The model Hamiltonian and the NEGF approach are explained in detail in Sec.\ \ref{sec:hm}. The numerical results and discussions are presented in Sec.\ \ref{sec:res}. Finally, the conclusions are summarized in Sec.\ \ref{sec:con}.

\begin{figure}[tp]
\begin{center}
\includegraphics[clip = true, width =\columnwidth]{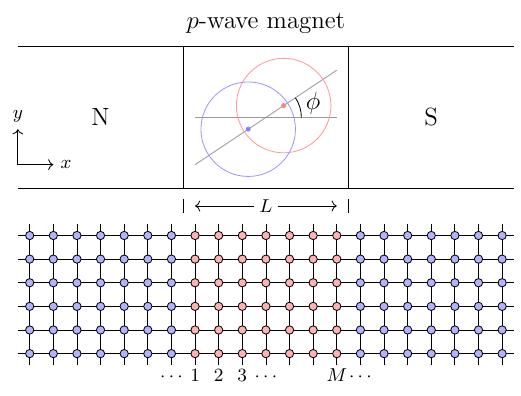}
\end{center}
\caption{(Top panel) Schematic image of the two-dimensional normal metal/$p$-wave magnet/superconductor junction. The longitudinal direction of the junction is along the $x$ axis. The central $p$-wave magnet region has a length $L$ and is located at $0<x<L$. The semi-infinite normal lead (N) and superconducting lead (S) are located at $x<0$ and $x>L$, respectively. In the central region, the Fermi surfaces for the spin-up and -down electron states are denoted by the red and blue circles, respectively. The angle between the line connecting the centers of the spin-polarized Fermi surfaces and the $x$-axis is represented by $\phi$. (Bottom panel) The two-dimensional square lattice on which the junction in (a) are discretized.
}
\label{fig:ju}
\end{figure}

\section{Model and formalism}
\label{sec:hm}

The two-dimensional (2D) tunnel junction under consideration is shown in Fig.\ \ref{fig:ju} (top panel), where the longitudinal direction of the junction is along the $x$ axis. The unconventional $p$-wave magnet is located at $0<x<L$, while the semi-infinite normal lead (N) and the superconducting lead (S) are located at $x<0$ and $x>L$, respectively. In terms of the Nambu spinor $\hat{\psi}_{\bm k}=(\hat{c}_{\bm k,\uparrow},\hat{c}_{\bm k,\downarrow},\hat{c}^\dagger_{-\bm k,\uparrow},\hat{c}^\dagger_{-\bm k,\downarrow})^T$, the system can be described by the Bogoliubov-de Gennes (BdG) Hamiltonian \cite{de2018superconductivity,PhysRevLett.97.067007} $\mathcal{H}_{BdG}=\frac{1}{2}\sum_{\bm k}\hat{\psi}_{\bm k}^\dagger\hat{\mathcal{H}}_{\bm k}\hat{\psi}_{\bm k}$ with
\begin{align}
     \hat{\mathcal{H}}_{\bm k}=\begin{pmatrix}
         \hat{h}(\bm k)&\hat{\Delta}(\bm k)\\
         \hat{\Delta}^\dagger(\bm k)&-\hat{h}^*(-\bm k)
     \end{pmatrix},\label{eq:bdg}
 \end{align} 
where $\bm{k}=(k_x,k_y)$ is the wave vector in the $x-y$ plane. $\hat\Delta(\bm k)=i\hat{s}_y\tilde{\Delta}$ is the conventional $s$-wave pair potential in the right superconducting lead, where $\hat{s}_y$ is the Pauli matrix in spin space, and $\tilde{\Delta}=\Delta\Theta(x-L)$ is the amplitude of the pair potential with $\Theta(x)$ being the Heaviside step function. The single-particle Hamiltonian for the system in the normal state takes the form \cite{PhysRevB.111.144508}
\begin{align}
    \hat{h}(\bm k)=\xi_{\bm k}\hat{s}_0+\tilde{\mathcal{J}}(k_x\cos\phi+k_y\sin\phi)\hat{s}_z,\label{eq:nor}
\end{align}
where $\xi_{\bm k}=\hbar^2\bm{k}^2/2m-\tilde{\mu}$ is the kinetic energy, $\tilde{\mu}=\mu+\delta\mu\Theta(x-L)$ is the chemical potential with $\delta\mu$ being the difference of the chemical potential between the superconducting and normal regions, and $\hat{s}_0$ is the identity matrix. $\tilde{\mathcal{J}}=\mathcal{J}\Theta(x)\Theta(L-x)$ describes the exchange energy of the $p$-wave magnet in the central region with $\mathcal{J}$ being the exchange coupling strength. This $p$-wave magnetism gives rise to two spin-polarized Fermi surfaces in the central region of the junction, which are shifted in the momentum space. The direction of the Fermi surface splitting is denoted by $\phi$, representing the angle between the line connecting the centers of the spin-polarized Fermi surfaces and the $x$ axis; see Fig.\ \ref{fig:ju} (top panel).

We employ the NEGF method \cite{PhysRevB.57.10972,PhysRevB.63.052512,PhysRevB.111.184504} to calculate the transverse tunneling Hall conductance. The continuum Hamiltonian is discretized on a 2D square lattice with lattice constant $a$, as shown in Fig.\ \ref{fig:ju} (bottom panel). The normal-state Hamiltonian is written on the discrete lattice as
\begin{align}
    H_N=\sum_{j}(\hat{c}^\dagger_{j}\check{h}_0\hat{c}_{j}+\hat{c}^\dagger_{j}\check{t}_x\hat{c}_{j+\delta_x}+\hat{c}^\dagger_{j}\check{t}_y\hat{c}_{j+\delta_y}+h.c.),\label{eq:lat}
\end{align}
where the index $j=(j_x,j_y)$ denotes the position of sites on the lattice, $\delta_{x(y)}$ is the lattice spacing between nearest neighbors along the $x$ ($y$) direction (with $\delta_x=\delta_y=a$), and $\hat{c}_{j}=(\hat{c}_{j,\uparrow},\hat{c}_{j,\downarrow})^T$ is the spinor operator acting on the spin space at site $j$. The $p$-wave magnet in the central region extends over $0\leq j_x\leq M$, with a length $L=Ma$. The on-site energy is $\check{h}_0=(-\mu+4t)\hat{s}_0$ with $t=\hbar^2/2ma^2$. $\check{t}_x=-t\hat{s}_0-i\tilde{\mathcal{J}}(\cos\phi/2a)\hat{s}_z$ and $\check{t}_y=-t\hat{s}_0-i\tilde{\mathcal{J}}(\sin\phi/2a)\hat{s}_z$ are the hopping energy between the nearest neighbor sites along the $x$ and $y$ directions, respectively.

We consider the periodic boundaries along the $y$ direction, so that $k_y$ is a good quantum number. Consequently, the model can be mapped onto a $k_y$-dependent one-dimensional squire lattice, where the discrete Hamiltonian in the BdG representation reads
\begin{align}
     H=\sum_{j_x,k_y}(\hat{\psi}^\dagger_{j_x}\check{H}_{0}\hat{\psi}_{j_x}+\hat{\psi}^\dagger_{j_x}\check{T}_{x}\hat{\psi}_{j_x+1}+h.c.).\label{eq:kh}
\end{align} 
Here $\check{H}_{0}$ represents the $k_y$-contracted on-site energy matrix at the $(j_x)_{\mathrm{th}}$ layer
\begin{align}
    \check{H}_{0}&=[-2t\cos(k_ya)+\check{h}_0]\hat{\tau}_z\hat{s}_0-\tilde{\Delta}\hat{\tau}_y\hat{s}_y\nonumber\\&\quad+\frac{\tilde{\mathcal{J}}\sin\phi\sin(k_ya)}{a}\hat{\tau}_0\hat{s}_z,
\end{align}
where $\hat{\tau}_i$ ($\hat{s}_i$) with $i=x,y,z$ is the Pauli matrix in Nambu (spin) space and $\hat{\tau}_0$ ($\hat{s}_0$) is the $2\times2$ identity matrix. $\check{T}_{x}$ denotes the coupling between the $(j_x)_{\mathrm{th}}$ and $(j_x+1)_{\mathrm{th}}$ layers, which is given by
\begin{align}
    \check{T}_{x}=-t\hat{\tau}_z\hat{s}_0-\frac{i\tilde{\mathcal{J}}\cos\phi}{2a}\hat{\tau}_0\hat{s}_z.
\end{align}
$\hat{\psi}_{j_x}$ is the shorthand notation of $\hat{\psi}_{j_x,k_y}$, which denotes the $y$-direction Fourier component of the Nambu spinor at site $j=(j_x,j_y)$

To evaluate the transverse current, we first derive the current operator by using the continuity equation, which in the discrete representation reads \cite{Csd}
\begin{align}
    \partial_t\hat{\rho}_{j}+\hat{\bm J}_{j}-\hat{\bm J}_{j-\delta}=0.
\end{align}
Here $\hat{\bm J}_{j}=(\hat{J}^x_{j},\hat{J}^y_{j})$ is the operator of the current density vector at site $j=(j_x,j_y)$, and $\delta=(\delta_x,\delta_y)$ is the lattice spacing between the nearest neighbors. The charge density operator at site $j$ is given by $\hat{\rho}_{j}=e\sum_{s=\uparrow,\downarrow}\hat{c}^\dagger_{j,s} \hat{c}_{j,s}$, which obeys the Heisenberg equation of motion $\partial_t\hat{\rho}_{j}=[\hat{\rho}_{j},H]/i\hbar$. The transverse current operator is obtained after completing the calculation of the commutator, which is given by
\begin{align}
    \hat{J}^y_{j}=\frac{ie}{\hbar}(\hat{c}^\dagger_{j}\check{t}_{y}\hat{c}_{j+1}-\hat{c}^\dagger_{j+1}\check{t}_{y}^\dagger\hat{c}_{j}),\label{eq:tj}
\end{align}
with $\check{t}_{y}$ representing the nearest-neighbor hopping matrix as specified in Eq.\ (\ref{eq:lat})

In a stationary state, the current can be calculated at any point in the system \cite{Csd}. Without loss of generality, we focus on the transverse current at the boundary of the normal lead ($j_x=0$). With the help of Eq.\ (\ref{eq:tj}), the transverse current is given by
\begin{align}
    J^y=\sum_{j_y}\langle\hat{J}^y_{j=(0,j_y)}\rangle=\frac{e}{h}\sum_{k_y}\int dE\,\mathrm{Tr}[\mathcal{M}G^{<}_{ee}],\label{eq:jy}
\end{align}
where $\mathcal{M}=2i\mathrm{Im}[\check{t}_{y}e^{ik_ya}]$, $G^{<}$ is the lesser Green's function in the energy domain, and the subscript $ee$ indicates the sum over the electron component in the Nambu space. $G^{<}$ can be expressed in terms of the self-energies via the Keldysh equation
\begin{align}
    G^<=G^r\Sigma^<G^a=G^r(\Sigma^<_L+\Sigma^<_R)G^a,\label{eq:kel}
\end{align}
where $\Sigma^<_{L,R}$ are the lesser self-energies coupling to the left normal lead and right superconducting lead, respectively. The retarded and advanced Green's functions read
\begin{align}
    G^{r}=[G^{a}]^\dagger=\frac{1}{E+i\eta-H_C-\Sigma_L^r-\Sigma_R^r},
\end{align}
where $E$ is the energy, $\eta\rightarrow0^+$, and $H_C$ is the Hamiltonian of the central region. The retarded self-energy $\Sigma_{L,R}^r$ due to coupling with the left and right leads can be calculated numerically by the recursive iteration method \cite{MPLopezSancho_1985,MPLopezSancho_1984}. By applying the Keldysh equation in Eq.\ (\ref{eq:kel}), Eq.\ (\ref{eq:jy}) can be further simplified to $J^y=\sum_{s=\uparrow,\downarrow}J^y_s$ with 
\begin{align}
    J^y_s&=\frac{ie}{h}\sum_{k_y}\int dE\,\mathrm{Tr}[\mathcal{M}_{s}(G^{r}_{eh}\Gamma_{L,h}G^{a}_{he})_{ss}(\bar{f}_L-f_L)\nonumber\\
    &\quad+\mathcal{M}_{s}(G^r\Gamma_RG^a)_{ee,ss}(f_0-f_L)],\label{eq:jssj}
\end{align}
where the subscripts $e/h$ and $s$ refer to the electron/hole component in the Nambu $e$-$h$ space and the spin component in the spin space, respectively. $f_0=f(E)$ is the Fermi distribution function. The shorthand notations $f_L$, $\bar{f}_L$ are $f_L=f(E-eV)$ and $\bar{f}_L=f(E+eV)$ with $V$ being the bias voltage. The line-width functions are given by $\Gamma_{L,R}=-2\mathrm{Im}\Sigma_{L,R}^r$. The corresponding transverse conductance for the spin-$s$ channel, calculated as $\partial J^y_s/\partial V$, is finally obtained as (see Appendix for details)
\begin{align}
        \sigma^{yx}_s&=\frac{e^2}{h}\mathrm{Im}\sum_{k_y} \mathcal{M}_{s}[(G^{r}_{he}\Gamma_{L,e}G^{a}_{eh})_{ss}+(G^{r}_{eh}\Gamma_{L,h}G^{a}_{he})_{ss}\nonumber\\
        &\quad+(G^r\Gamma_RG^a)_{ee,ss}].\label{eq:tc}
\end{align}
The transverse conductance can be decomposed into three distinct components: the first two terms in Eq.\ (\ref{eq:tc}) originate from Andreev reflection, while the third term arises from quasiparticle tunneling across the junction.

\begin{figure}[tp]
\begin{center}
\includegraphics[clip = true, width =\columnwidth]{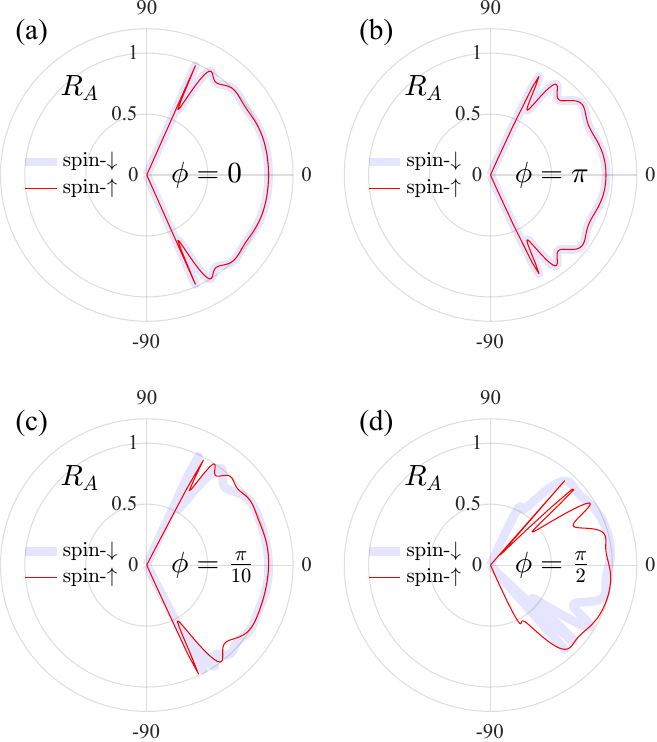}
\end{center}
\caption{Zero-bias Andreev reflection probabilities versus the transverse momentum $k_y$ ($-\pi<k_y<\pi$) for $\phi=0$ (a), $\phi=\pi$ (b) $\phi=\pi/10$ (c) and $\phi=\pi/2$ (d). The differences of the chemical potential are set as $\delta\mu=0$ [(a), (c) and (d)] and $\delta\mu=0.5t$ (b). The other parameters are $M=10$, $\mu=1.2t$, $\mathcal{J}=0.6t$ and $\Delta=0.015t$. The Andreev reflection coefficients for spin-up and -down channels are denoted by red and blue lines, respectively.}
\label{fig:AR}
\end{figure}

\section{Results and discussions} 
\label{sec:res}
We begin by analyzing the Andreev reflection process in the left normal metal, which can aid in better understanding the tunneling spin Hall effect. Within the framework of NEGF formalism, the Andreev reflection coefficient can be obtained by $R_A=\mathrm{Tr}[\Gamma_{L,e}G^{r}_{eh}\Gamma_{L,h}G^{a}_{he}]$ \cite{PhysRevB.59.3831,PhysRevB.59.13126}. Without loss of generality, we set $a=1$ throughout our calculation and $t=\SI{1}{\eV}$ as the energy unit. Since the $p$-wave magnetism does not couple different spin channels, the spin-flip Andreev process is always absent. The zero-bias spin-conserving Andreev reflection probabilities as a function of the transverse momentum $k_y$ ($-\pi<k_y<\pi$) are shown in Fig.\ \ref{fig:AR} at $\mu=1.5t$ and $\mathcal{J}=0.6t$. Although the $p$-wave magnetism shifts the Fermi surfaces of the electrons with opposite spins, the two spin channels are indistinguishable in terms of their Andreev reflection properties when the shifts are along the $x$ axis, \textit{i.e.}, $\phi=0$ or $\phi=\pi$, where the line connecting the centers of the spin-split Fermi surface is parallel to the normal direction of the junction; see Figs.\ \ref{fig:AR}(a) and \ref{fig:AR}(b). For $\delta\mu=0$, in which the Fermi energies of the normal and superconducting regions are equal, the Andreev reflection is almost insensitive to the $p$-wave magnetism over a wide range of the transverse momenta. The Andreev reflection exhibits the similar behavior to that of the Andreev reflection in traditional normal metal/superconductor junctions with a transparent interface, where the Andreev reflection probability is unit in the subgap regime \cite{Kashiwaya_2000}; see Fig.\ \ref{fig:AR}(a). This perfect Andreev reflection is suppressed in the oblique incident regime due to the Fermi surface mismatch caused by the $p$-wave magnetism at large transverse momenta. The similar behavior of the Andreev reflections also exists at $\phi=\pi$, as shown in Fig.\ \ref{fig:AR}(b), where $\delta\mu=0.5t$. The perfect Andreev reflection with $R_A=1$ is absent even at the normal incidence due to the mismatch of the Fermi wave vector caused by nonzero $\delta\mu$.

A spin-resolved Andreev reflection process arises when the line connecting the centers of the spin-split Fermi surfaces deviates from the normal direction of the junction, \textit{i.e.}, $\phi\neq0$ or $\pi$. The $k_y$-dependent Andreev reflection probabilities are presented in Fig.\ \ref{fig:AR}(c) at $\phi=\pi/10$ and $\delta\mu=0$. It is shown that the spin-resolved Andreev reflection coefficients remain nearly identical over a wide range of angles close to the normal incidence, rendering the opposite spin channels effectively indistinguishable. However, at oblique incidence with large $k_y$, the Andreev reflection behavior becomes strongly spin-dependent, giving rise to a pronounced spin contrast. With increasing $\phi$, the spin-resolved nature of the Andreev reflection is markedly enhanced, indicating a strong transverse-momentum-dependence of the spin-selective reflection processes, as shown in Fig.\ \ref{fig:AR}(d) where $\phi=\pi/2$. The spin-up electron has a large Andreev reflection probability for the transverse momentum in the range $-\pi/2\leq k_y\leq0$, leading to the skew Andreev reflection, \textit{i.e.}, $R_A^s(k_y)\neq R_A^s(-k_y)$; see the red line in Fig.\ \ref{fig:AR}(d). The spin-down electron exhibits a similarly asymmetric behavior but is skewed in the opposite direction [blue line in Fig.\ \ref{fig:AR}(d)]. Since the unconventional $p$-wave magnet preserves time-reversal symmetry, the Andreev reflection probabilities for electrons with opposite spins are symmetric, \textit{i.e.}, $R_A^s(k_y)\neq R_A^{\bar{s}}(-k_y)$, where $\bar{s}=-s$. Consequently, the total Andreev reflection probability ($R^s_A+R^{\bar{s}}_A$) remains even with respect to the transverse momentum.

\begin{figure}[tp]
\begin{center}
\includegraphics[clip = true, width =0.85\columnwidth]{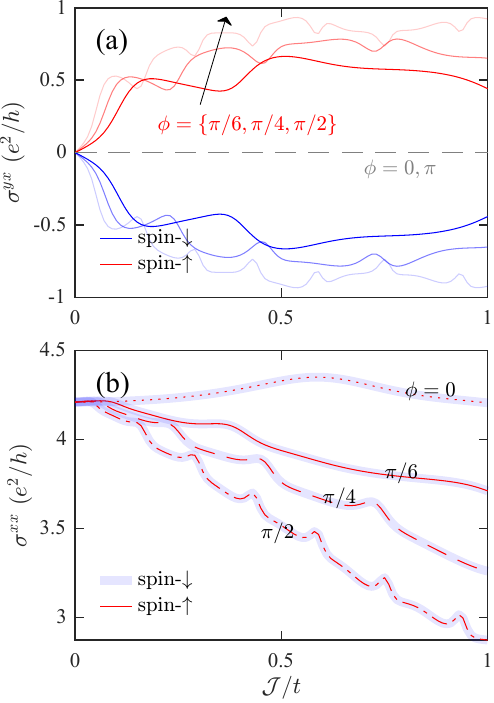}
\end{center}
\caption{The transverse conductance (a) and longitudinal conductance (b) versus the $p$-wave exchange coupling strength $\mathcal{J}$ for spin-up and -down channels, which are denoted by red and blue lines, respectively. The numerical parameters are $M=10$, $E=0$, $\mu=1.2t$, $\delta\mu=0.3t$, $\mathcal{J}=0.6t$ and $\Delta=0.018t$.}
\label{fig:cd}
\end{figure}

As a consequence of the spin-contrasting asymmetric Andreev reflection, the carriers with opposite spins are deflected into different transverse directions, leading to a net transverse spin current. The transverse conductance $\sigma^{yx}$ for each spin channel as a function of the $p$-wave magnet strength $\mathcal{J}$ is numerically calculated from Eq.\ (\ref{eq:tc}) and is shown in Fig.\ \ref{fig:cd}(a). The transverse current is absent when the line connecting the centers of the spin-split Fermi surfaces is parallel to the normal direction of the junction at $\phi=0$ or $\pi$, as the gray dashed line shown in Fig.\ \ref{fig:cd}(a), where the transverse conductance is zero. The absence of $\sigma^{yx}$ is attributed to the spin-degenerate and symmetric nature of the Andreev reflection discussed earlier [see Figs.\ (\ref{fig:AR})(a) and (\ref{fig:AR})(b)]. However, a finite spin-contrasting transverse current appears for nonzero $\phi$. We present the spin-dependent transverse conductance for $\phi=\pi/6$, $\pi/4$ and $\pi/2$ in Fig.\ \ref{fig:cd}(a). It is shown that the presence of the unconventional $p$-wave magnet ($\mathcal{J}>0$) gives rise to a finite transverse conductance, which exhibits an overall increasing trend with increasing $\mathcal{J}$. For a given $\phi$, the transverse conductance contributions from opposite spin channels are equal in magnitude but opposite in sign, \textit{i.e.}, $\sigma^{yx}_\uparrow=-\sigma^{yx}_\downarrow$. This quantitative relation of the transverse conductance between different spin channels is protected by the time-reversal symmetry, which indicates that the carriers with opposite spins tend to propagate in opposite transverse directions, leading to a finite transverse spin current $J^y_{\mathrm{spin}}=(\hbar/2e)(J^y_{\uparrow}-J^y_{\downarrow})$ and the transverse spin Hall conductance $\sigma^{yx}_{\mathrm{spin}}=(\hbar/e)\sigma^{yx}_\uparrow$. The net transverse charge conductance is always absent. It should also be noted that as $\phi$ varies from $0$ to $\pi/2$, the transverse conductance is generally enhanced at the larger $\phi$. This behavior originates from the increased parallel component of $\mathcal{J}$ at larger $\phi$, which in turn intensifies the asymmetry between opposite-spin Andreev reflection channels; see Figs. \ref{fig:AR}(c) and \ref{fig:AR}(d).

\begin{figure}[tp]
\begin{center}
\includegraphics[clip = true, width =0.85\columnwidth]{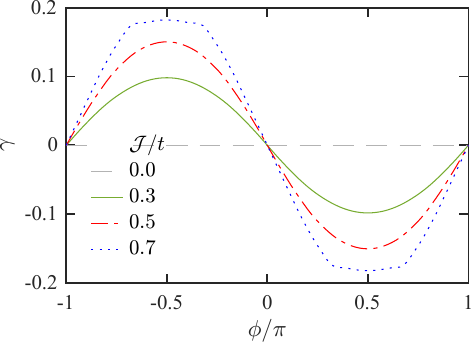}
\end{center}
\caption{The spin Hall angle $\gamma$ as a function of $\phi$. The numerical parameters are $M=10$, $E=0$, $\mu=3.2t$, $\delta\mu=0$ and $\Delta=0.018t$.}
\label{fig:sh}
\end{figure}

The longitudinal charge conductance can be calculated by the Blonder-Tinkham-Klapwijk formula \cite{PhysRevB.25.4515} $\sigma^{xx}=(e^2/h)\sum_{k_y}(T+2R_A)$, where $T=\mathrm{Tr}[\Gamma_{L,e}(G^r\Gamma_RG^a)_{ee}]$ is the probability for the quasiparticles tunneling. For different spin channels, the longitudinal conductance exhibits the same behavior as a function of the $p$-wave magnet strength $\mathcal{J}$, as shown in Fig.\ \ref{fig:cd}(b). In contrast to the transverse conductance, the longitudinal conductance decreases with increasing $\mathcal{J}$ or $\phi$, owing to the stronger mismatch between the Fermi momenta of the left normal lead and the central $p$-wave magnet region. The efficiency of the charge-to-spin conversion is characterized by the spin Hall angle $\gamma$, which is given by $\gamma=(2e/\hbar)\times\sigma^{yx}_{\mathrm{spin}}/\sigma^{xx}$. $\gamma$ as a function of $\phi$ is presented in Fig.\ \ref{fig:sh}. The nonzero $\gamma$ appears when the line connecting the centers of the spin-split Fermi surfaces deviates from the normal direction of the junction ($\phi\neq0$ or $\pi$) and reaches its maximum value at $\phi=\pm\pi/2$. The odd dependence of $\gamma$ on $\phi$ is attributed to the rotational symmetry of the $p$-wave magnet in the momentum space. Under $\phi\rightarrow-\phi$, the Fermi surfaces associated with different spin indices are exchanged, leading to a reversal of the transverse spin current and thus a sign change in $\gamma$. We note that the spin Hall angles in conventional semiconductors are typically very small (on the order of $0.0001$ to $0.001$) \cite{doi:10.1126/science.1105514,ando2012observation}. In contrast, significant enhancement can be achieved in graphene via single-impurity scattering, resulting in spin Hall angles ranging from $0.01$ to $0.1$ \cite{PhysRevLett.112.066601}. Notably, $\gamma$ can reach the order of $0.2$ in our model, which is comparable to the large spin Hall angle obtained in graphene grown by chemical vapour deposition \cite{balakrishnan2014giant}, indicating that a highly efficient charge-to-spin conversion can be generated via the Andreev reflection processes mediated by the $p$-wave magnetism.

The underlying physics of the predicted tunneling spin Hall effect in our model is the same as that of the anomalous Hall effect previously reported in the ferromagnetic tunnel junctions \cite{PhysRevB.100.060507,PhysRevB.111.054512}, both arising from the spin-dependent momentum filtering during the electron tunneling process. However, the origins of the momentum filtering for these effects are different. In the latter case, the effect is attributed to the spin-dependent anisotropic scattering caused by the spin-orbit coupling potential at the ferromagnet/superconductor interface. In contrast, the momentum filtering in our model arises from the mismatch between the Fermi surfaces of the electrode and the central region of the junction, where the spin-dependent Fermi surfaces are shifted due to the $p$-wave magnetism. In fact, the splitting of the spin-dependent Fermi surfaces in our model is reminiscent of that in tilted Dirac systems \cite{PhysRevB.110.024511,PhysRevLett.131.246301}, where the Dirac cones tilt in opposite directions for different valleys, leading to valley-dependent shifts of the Fermi surfaces in momentum space and thereby giving rise to the tunneling valley Hall effect. However, in contrast to the tilt-induced tunneling valley Hall effect, where a tilt-driven transverse charge current can arise despite preserved time-reversal symmetry, the transverse charge current is always absent in our model.

\section{Conclusions} 
\label{sec:con}

To conclude, we theoretically study the transverse charge and spin transport in the normal metal/$p$-wave magnet/superconductor junction using the nonequilibrium Green's function formalism. It is shown that the $p$-wave magnetism may induce a spin-contrasting Andreev reflection in the normal lead. Notably, this spin-resolved Andreev reflection probability is strongly asymmetric with respect to the transverse momentum, giving rise to the spin-dependent transverse transport, where the carriers with opposite spins propagate along opposite transverse directions of the tunnel junction, resulting in a finite transverse pure spin current with zero net charge. We further demonstrate that the transverse spin conductance is determined by the direction of Fermi surface splitting in the $p$-wave magnet. A large spin Hall angle occurs when the line connecting the centers of the spin-split Fermi surfaces is perpendicular to the normal direction of the junction, indicating a highly efficient charge-to-spin conversion.

\onecolumngrid

\appendix

\section{Derivation of the transverse conductance}

In this appendix, we provide a detailed derivation of the transverse conductance using the NEGF formalism on a 2D discrete lattice. With the aid of the Keldysh equation
\begin{align}
    G^<=G^r\Sigma^<G^a=G^r(\Sigma^<_L+\Sigma^<_R)G^a=G^r(i\mathfrak{f}_L\Gamma_L+if_0\Gamma_R)G^a,
\end{align}
where the shorthand notation $\mathfrak{f}_L=\mathrm{diag}[f_L,\bar{f}_L]$ with $f_L$, $\bar{f}_L$, and $f_0$ being the Fermi distribution functions as specified in Eq.\ (\ref{eq:jssj}), the transverse current in Eq.\ (\ref{eq:jy}) can be reduced to 
\begin{align}
    J^y&=\frac{e}{h}\sum_{k_y}\int dE\,\mathrm{Tr}[\mathcal{M}G^{<}_{ee}]\nonumber\\
    &=\frac{ie}{h}\sum_{k_y}\int dE\,\mathrm{Tr}[\mathcal{M}(G^r\mathfrak{f}_L\Gamma_LG^a)_{ee}+\mathcal{M}(G^rf_0\Gamma_RG^a)_{ee}]\nonumber\\
    &=\frac{ie}{h}\sum_{k_y}\int dE\,\mathrm{Tr}[\mathcal{M}G^{r}_{ee}\Gamma_{L,e}G^{a}_{ee}f_L+\mathcal{M}G^{r}_{eh}\Gamma_{L,h}G^{a}_{he}\bar{f}_L+\mathcal{M}(G^r\Gamma_RG^a)_{ee}f_0].\label{eq:sscc}
\end{align}
where the subscript $e$ ($h$) denotes the electron (hole) component in Nambu electron-hole space. $\mathcal{M}$ characterizes the behavior of the $y$-directional group velocity of the electrons in the left normal lead, which is given by 
\begin{align}
    \mathcal{M}=2i\mathrm{Im}[\check{t}_{y}e^{ik_ya}]=-2it\sin(k_ya)\hat{s}_0.\label{eq:xi}
\end{align}
Using the identity for the spectral function $A=i(G^r-G^a)=G^r\Gamma G^a=G^r(\Gamma_L+\Gamma_R) G^a$, one obtains that 
\begin{align}
    A_{ee}=G^{r}_{ee}\Gamma_{L,e}G^a_{ee}+G^{r}_{eh}\Gamma_{L,h} G^a_{he}+(G^{r}\Gamma_R G^a)_{ee},\label{eq:sp}
\end{align}
Eq.\ (\ref{eq:sp}) can help us further simplify Eq.\ (\ref{eq:sscc}) as follows
\begin{align}
    J^y&=\frac{ie}{h}\sum_{k_y}\int dE\,\mathrm{Tr}[\mathcal{M}(A_{ee}-G^{r}_{eh}\Gamma_{L,h} G^a_{he}-(G^{r}\Gamma_R G^a)_{ee})f_L+\mathcal{M}G^{r}_{eh}\Gamma_{L,h}G^{a}_{he}\bar{f}_L+\mathcal{M}(G^r\Gamma_RG^a)_{ee}f_0]\nonumber\\
    &=\frac{ie}{h}\sum_{k_y}\int dE\,\mathrm{Tr}[\mathcal{M}G^{r}_{eh}\Gamma_{L,h}G^{a}_{he}(\bar{f}_L-f_L)+\mathcal{M}(G^r\Gamma_RG^a)_{ee}(f_0-f_L)+\mathcal{M}A_{ee}f_L].\label{eq:yy1}
\end{align}
In equilibrium, $f_L=\bar{f}_L=f_0$, the integrand in Eq.\ (\ref{eq:yy1}) becomes $\mathrm{Tr}[\mathcal{M}A_{ee}f_L]$. We note that $\mathcal{M}$ is an odd function of the transverse momentum $k_y$ [see Eq.\ (\ref{eq:xi})], whereas the spectral function $A_{ee}$ (which plays the role of a generalized density of states in the normal lead) is an even function of the transverse momentum $k_y$. Consequently, the summation of $\mathrm{Tr}[\mathcal{M}A_{ee}f_L]$ over a symmetric interval in $k_y$ (namely, the interval $-\pi<k_y<\pi$) vanishes, giving rise to the absence of the transverse current at equilibrium. The transverse current caused by the tunneling process is given by
\begin{align}
     J^y=\frac{ie}{h}\sum_{k_y}\int dE\,\mathrm{Tr}[\mathcal{M}G^{r}_{eh}\Gamma_{L,h}G^{a}_{he}(\bar{f}_L-f_L)+\mathcal{M}(G^r\Gamma_RG^a)_{ee}(f_0-f_L)].\label{eq:jy5}
 \end{align} 
It should be noted that $\mathcal{M}$ is a diagonal matrix in spin space [see Eq.\ (\ref{eq:xi})]. Therefore, the trace in Eq.\ (\ref{eq:jy5}) can be decomposed into a sum of traces over individual spin components, \textit{i.e.}, $J^y=\sum_{s=\uparrow,\downarrow} J^y_s$ with
\begin{align}
    J^y_s&=\frac{ie}{h}\sum_{k_y}\int dE\,\mathrm{Tr}[\mathcal{M}_s(G^{r}_{eh}\Gamma_{L,h}G^{a}_{he})_{ss}(\bar{f}_L-f_L)+\mathcal{M}_{s}(G^r\Gamma_RG^a)_{ee,ss}(f_0-f_L)]\nonumber\\
    &=\frac{ie}{h}\sum_{k_y}\int dE\,\mathrm{Tr}[\mathcal{M}_{s}G^{r}_{eh,ss}\Gamma_{L,hs}G^{a}_{he,ss}(\bar{f}_L-f_L)+\mathcal{M}_{s}G^{r}_{eh,s\bar{s}}\Gamma_{L,h\bar{s}}G^{a}_{he,\bar{s}s}(\bar{f}_L-f_L)\nonumber\\&\quad+\mathcal{M}_{s}(G^r\Gamma_RG^a)_{ee,ss}(f_0-f_L)]\nonumber\\
    &=\frac{e}{h}\sum_{k_y}\int dE\,\kappa_{s}\mathrm{Tr}[\Gamma_{L,es}G^{r}_{eh,ss}\Gamma_{L,hs}G^{a}_{he,ss}(\bar{f}_L-f_L)+\Gamma_{L,es}G^{r}_{eh,s\bar{s}}\Gamma_{L,h\bar{s}}G^{a}_{he,\bar{s}s}(\bar{f}_L-f_L)\nonumber\\&\quad+\Gamma_{L,es}(G^r\Gamma_RG^a)_{ee,ss}(f_0-f_L)]\nonumber\\
    &=\frac{e}{h}\sum_{k_y}\int dE\,\kappa_{s}[R_A^{ss}(\bar{f}_L-f_L)+R_A^{\bar{s}s}(\bar{f}_L-f_L)+T^{s}(f_0-f_L)],\label{eq:jys}
\end{align}
where the dimensionless parameter $\kappa_s$ is given by 
\begin{align}
    \kappa_{s}=i\mathcal{M}_{s}/\Gamma_{L,es}.\label{eq:kpp}
\end{align}
$R_A^{ss}$, $R_A^{\bar{s}s}$, and $T^{s}$ in Eq.\ (\ref{eq:jys}) are the spin-conserving Andreev reflection probability, spin-flip Andreev reflection probability and the tunneling probability, which are given by
\begin{gather}
    R_A^{ss}=\mathrm{Tr}[\Gamma_{L,es}G^{r}_{eh,ss}\Gamma_{L,hs}G^{a}_{he,ss}],\quad R_A^{\bar{s}s}=\mathrm{Tr}[\Gamma_{L,es}G^{r}_{eh,s\bar{s}}\Gamma_{L,h\bar{s}}G^{a}_{he,\bar{s}s}],\quad T^{s}=\mathrm{Tr}[\Gamma_{L,es}(G^r\Gamma_RG^a)_{ee,ss}],\label{eq:AR}
\end{gather}
respectively. Since the trace is taken over scalar quantities rather than matrix-valued ones, the trace operation can be omitted in Eq.\ (\ref{eq:AR}). Notice that the integration limits extend over the entire energy range (from $-\infty$ to $\infty$) and that the Fermi distribution function satisfies $\bar{f}_L(-E)=1-f_L(E)$. Therefore, by changing the integration variable $E\rightarrow-E$ in the first two terms of Eq.\ (\ref{eq:jys}), we further obtain 
\begin{align}
    J^y_s&=\frac{e}{h}\sum_{k_y}\int dE\,([\kappa_{s}R_A^{s}]_{(-E)}+[\kappa_{s}(T^s+R_A^{s})]_{(+E)})(f_0-f_L)\nonumber\\
    &=\frac{e}{h}\sum_{k_y}\int dE\,(\bar{\kappa}_{s}\bar{R}_A^{s}+\kappa_{s}R_A^{s}+\kappa_{s}T^s)(f_0-f_L)\nonumber\\
    &=-\frac{e}{h}\sum_{k_y}\int dE\,(\bar{\kappa}_{s}\bar{R}_A^{s}+\kappa_{s}R_A^{s}+\kappa_{s}T^s)(f_L-f_0),\label{eq:tps}
\end{align}
where $R^s_A=R^{ss}_A+R^{\bar{s}s}_A$ is the total Andreev reflection probability. For simplify, we use the shorthand notation $\bar{\kappa}_{s}$ and $\bar{R}_A^s$ represent $\kappa_{s}(-E)$ and $R_A^s(-E)$, respectively.

The transverse conductance induced by the tunneling process can be obtained by $\partial J^y/\partial V$. At zero temperature, the Fermi distribution function satisfies $\partial f(E)/\partial E=-\delta(E)$. With the help of Eqs.\ (\ref{eq:kpp}-\ref{eq:tps}), the zero-temperature transverse conductance is given by
\begin{align}
    \sigma^{yx}&=-\frac{e^2}{h}\sum_{k_y,s}(\bar{\kappa}_{s}\bar{R}_A^{s}+\kappa_{s}R_A^{s}+\kappa_{s}T^s)\nonumber\\
    &=\frac{e^2}{h}\mathrm{Im}\sum_{k_y,s} \mathcal{M}_s[(G^{r}_{he}\Gamma_{L,e}G^{a}_{eh})_{ss}+(G^{r}_{eh}\Gamma_{L,h}G^{a}_{he})_{ss}+(G^r\Gamma_RG^a)_{ee,ss}].\label{eq:tcct}
\end{align}

\section{Generalized to continuum models}

To generalize the transverse conductance formula in Eq.\ (\ref{eq:tcct}) to the continuum model, we start by finding the expression for $\kappa_s$ and $\bar{\kappa}_s$ in the continuum limit. 

For the left normal lead, the Hamiltonian can be mapped onto a $k_y$-dependent one-dimensional lattice since we consider the periodic boundaries along the $y$ direction, which can be written as $H_L=\sum_{k_y,s}h_{k_y,s}$ with
\begin{align}
    h_{k_y,s}=\sum_{j}(\varepsilon_{k_y}^sc^\dagger_{j,k_y,s}c_{j,k_y,s}-tc^\dagger_{j,k_y,s}c_{j+1,k_y,s}+h.c.).\label{eq:hks}
\end{align}
Here $\varepsilon_{k_y}^s=-\mu-4t-2t\cos(k_ya)$ is the $k_y$-dependent on-site energy. Eq.\ (\ref{eq:hks}) leads to the energy dispersion $E_{k_y}^s=\varepsilon_{k_y}^s-2t\cos(k^e_{s}a)$ with $k^e_{s}$ being the longitudinal wave vector for the electron states. The surface Green's function can be obtained by solving the tight-binding equations directly
\begin{gather}
    (E-\varepsilon_{k_y}^s)g^{r,e}_{n,1}+tg^{r,e}_{n-1,1}+tg^{r,e}_{n+1,1}=0,\quad (n>1),\label{eq:sllll}\\
    (E-\varepsilon_{k_y}^s)g^{r,e}_{1,1}+tg^{r,e}_{2,1}=1,\quad (n=1),\label{eq:sllll2}
\end{gather}
where $g^{r,e}$ is the retarded Green's function of the left normal lead. The energy dispersion together with Eq.\ (\ref{eq:sllll}) leads to 
\begin{align}
    g^{r,e}_{n,1}=g^{r,e}_{1,1}e^{ik^e_s(n-1)a}.\label{eq:g11}
\end{align}
By substituting Eq.\ (\ref{eq:g11}) into Eq.\ (\ref{eq:sllll2}), the electron component of the surface Green's function for the spin-$s$ channel is obtained 
\begin{align}
    g_{L,es}=g^{r,e}_{1,1}=(E-\varepsilon_{k_y}^s-te^{ik^e_sa})^{-1}=-\frac{1}{t}e^{ik^e_sa},
\end{align}
giving rise to the self-energy $\Sigma_{L,es}^{r}=t^2g_{L,es}=-te^{ik^e_sa}$. Consequently, the line-width function is obtained
\begin{align}
    \Gamma_{L,es}=i(\Sigma_{L,es}^{r}-\Sigma_{L,es}^{r\dagger})=2t\sin(k^e_s a).\label{eq:ge}
\end{align}
The hole components can be derived using the same procedure, which is given by
\begin{align}
    \Gamma_{L,hs}=-2t\sin(k^h_s a),\label{eq:gh}
\end{align}
with $k^h_s$ being the longitudinal wave vector for the hole states. 

Consequently, with the help of Eqs.\ (\ref{eq:xi}), (\ref{eq:kpp}), (\ref{eq:ge}) and (\ref{eq:gh}), in continuum limit ($a\rightarrow0$), we obtain 
\begin{gather}
\kappa_s(E)=\frac{2t\sin(k_ya)}{\Gamma_{L,es}(E)}=\frac{\sin(k_ya)}{\sin(k^e_s a)}\simeq\frac{k_y}{k^e_s},\\
    \bar{\kappa}_s=\kappa_s(-E)=\frac{2t\sin(k_ya)}{\Gamma_{L,es}(-E)}=\frac{2t\sin(k_ya)}{\Gamma_{L,hs}(E)}=-\frac{\sin(k_ya)}{\sin(k^h_s a)}\simeq-\frac{k_y}{k^h_s}.
\end{gather}
The current conservation requires $R^{ss}+R^{s\bar{s}}+R_A^{ss}+R_A^{s\bar{s}}+T^s=1$, where $R^{ss}$ and $R^{s\bar{s}}$ are the spin-conserving and spin-flip normal reflection probabilities, respectively. The transverse current in Eq.\ (\ref{eq:tps}) can be written as
\begin{align}
    J^y_s&=\frac{e}{h}\sum_{k_y,s}\int dE\,(\bar{\kappa}_{s}\bar{R}_A^{s}+\kappa_{s}R_A^{s}+\kappa_{s}T^s)(f_0-f_L)\nonumber\\
    &=\frac{e}{h}\sum_{q,s}\int dE\,(\bar{\kappa}_{s}\bar{R}_A^{ss}+\bar{\kappa}_{\bar s}\bar{R}_A^{s\bar{s}}-\kappa_{s}R^{ss}-\kappa_{\bar{s}}R^{s\bar{s}})(f_0-f_L),\nonumber\\
    &=\frac{e}{h}\sum_{k_y,s}\int dE\,(-\frac{k_y}{k^h_s}\bar{R}_A^{ss}-\frac{k_y}{k^h_{\bar{s}}}\bar{R}_A^{s\bar{s}}-\frac{k_y}{k^e_{s}}R^{ss}-\frac{k_y}{k^e_{\bar{s}}}R^{s\bar{s}})(f_0-f_L)\nonumber\\
    &=\frac{e}{h}\sum_{k_y,s}\int dE\,(\frac{k_y}{k^e_s}\frac{k^e_s}{k^h_s}\bar{R}_A^{ss}+\frac{k_y}{k^e_s}\frac{k^e_s}{k^h_{\bar{s}}}\bar{R}_A^{s\bar{s}}+\frac{k_y}{k^e_s}\frac{k^e_s}{k^e_{s}}R^{ss}+\frac{k_y}{k^e_s}\frac{k^e_s}{k^e_{\bar{s}}}R^{s\bar{s}})(f_L-f_0)\nonumber\\
    &=\frac{e}{h}\sum_{k_y,s}\int dE\,\frac{k_y}{k^e_s}(|\bar{r}_A^{ss}|^2+|\bar{r}_A^{s\bar{s}}|^2+|r^{ss}|^2+|r^{s\bar{s}}|^2)(f_L-f_0),
\end{align}
where $r^{ss'}_{(A)}$ is the corresponding reflection amplitude for the reflection probability $R^{ss'}_{(A)}$. Finally, the zero-temperature transverse conductance at $E=eV$ obtained by $\partial J^y/\partial V$ in continuum limit is given by
\begin{align}
    \sigma^{yx}(eV)=\frac{e^2}{h}\sum_{k_y,s}\frac{k_y}{k^e_s}(|r_A^{ss}(-eV)|^2+|r_A^{s\bar{s}}(-eV)|^2+|r^{ss}(eV)|^2+|r^{s\bar{s}}(eV)|^2),
\end{align}
which is extensively employed to evaluate the transverse conductance in continuum models \cite{PhysRevB.111.054512,PhysRevLett.115.056602,PhysRevB.110.024511}.

\twocolumngrid

\end{document}